\documentclass[preprint,showkeys,preprintnumbers,amsmath,amssymb]{revtex4}

\usepackage{graphicx}
\usepackage{dcolumn}
\usepackage{bm}
\usepackage{enumerate}

\usepackage{xcolor}

\DeclareMathOperator{\sign}{sign}

\begin{document}

\author{R. Rossi Jr.}
\email{romeu.rossi@ufv.br}
\affiliation{Universidade Federal de Vi{\c c}osa - Campus Florestal,
LMG818 Km6, Minas Gerais, Florestal 35690-000, Brazil}

\author{Leonardo A. M. Souza}
\email{leonardoamsouza@ufv.br}
\affiliation{Universidade Federal de Vi{\c c}osa - Campus Florestal,
LMG818 Km6, Minas Gerais, Florestal 35690-000, Brazil}

\date{\today}
\title{Tests of hidden variable models by the relaxation of the measurement independence condition}

\begin{abstract}
Bell inequalities or Bell-like experiments are supposed to test hidden variable theories based on three intuitive assumptions:  determinism, locality and measurement independence. If one of the assumptions of Bell inequality is properly relaxed, the probability distribution of the singlet state, for example, can be reproduced by a hidden variable model. Models that deal with the relaxation of some condition above, with more than one hidden variable, have been studied in the literature nowadays. In this work the relation between the number of hidden variables and the degree of relaxation necessary to reproduce the singlet correlations is investigated. For the examples studied, it is shown that the increase of the number of hidden variables does not allow for more efficiency in the reproduction of quantum correlations.
   
\end{abstract}


\maketitle

\section{Introduction}

One of the most intriguing features of quantum mechanics is that a multipartite state may present genuine (and intrinsic) quantum correlations. Different from classical mechanics, quantum correlations can not be described by a model simultaneously consistent with: determinism, locality and measurement independence \cite{art1,art2,art3,art4}. However, it is possible to reproduce quantum correlations in a model in which at least one of the previous assumptions is partially relaxed \cite{art5,art7,art8,art9,art10}. For instance, the reproduction of the singlet state correlations, by the relaxation of the measurement independence condition (MIC), was studied in Refs. \cite{art8,art9,art10}. Different models have been considered and it was shown in Reference \cite{art10} a model with the lowest degree of relaxation necessary to reproduce the statistics of the singlet. 

Acoording to the MIC, the measurement set variables ($x$ and $y$), in the context of Bell inequality scheme,  must be independent of the hidden variable. In the language of the theory of causal models \cite{pearl, spirtes}, this statement means that there must be no causal connections between the hidden variable and $x$ or $y$. Due to our lack of knowledge about the hidden variables, one can conceive models with a variety of them, interacting and affecting the values of the observable variables \cite{art11,art12}.

A question may arise: is it possible to use this freedom and consider a larger number of hidden variables to reproduce the singlet statistics, through the violation of the MIC, in a more efficient way? The degree of violation of the MIC, that measures the efficiency of a model to reproduce the singlet statistics, can be defined as the mutual information $I(\lambda:x,y)$ \cite{art13}. Compared to all models shown in the literature, the model of reference \cite{art10} is the most efficient, due to the relation among the measurement set variables considered in this reference.       

For models with more than one hidden variable the conditions for the violation of the measurement independence are different from the conditions of the traditional model (with one hidden variable). Here we consider models with three hidden variables ( that were presented in \cite{art11} and \cite{art12}). The violation of the MIC for these models is given when causal connections among the hidden variables are assumed.

In this work,  we investigate the possibility of increase the efficiency of a model (with hidden variables) in reproducing the singlet probability, through the violation of the MIC, by growing the number of hidden variables. We show that the efficiency of the model presented in \cite{art10} can not be increased just by adding hidden variables within our approach.

\section{Reproducing the Singlet Correlation}

Here we apply some tools developed in the theory of causal models \cite{pearl, spirtes} that are suitable for analysis of systems with hidden variables. This was also done in references \cite{Spekkens, rossi, chaves}. We consider three different causal models and study the reproduction of singlet state correlations through violation of the measurement independence condition. From now on, any variable $\lambda$ or $\lambda_i$ represents a different hidden variable. 

\subsection{The First Causal Model (Bell Scheme)}

In a traditional Bell experiment two subsystems, which may have interacted previously, are spatially separated and measured by two observers: Alice and Bob. The variable $x$ and $y$ are the setting variable, they describe the possible measurements that can be chosen by Alice and Bob, respectively. The variables $a$ ($b$) represents the possible outcome of measurements of $x$ ($y$).

\begin{figure}[h]
\centering
  \includegraphics[scale=0.5]{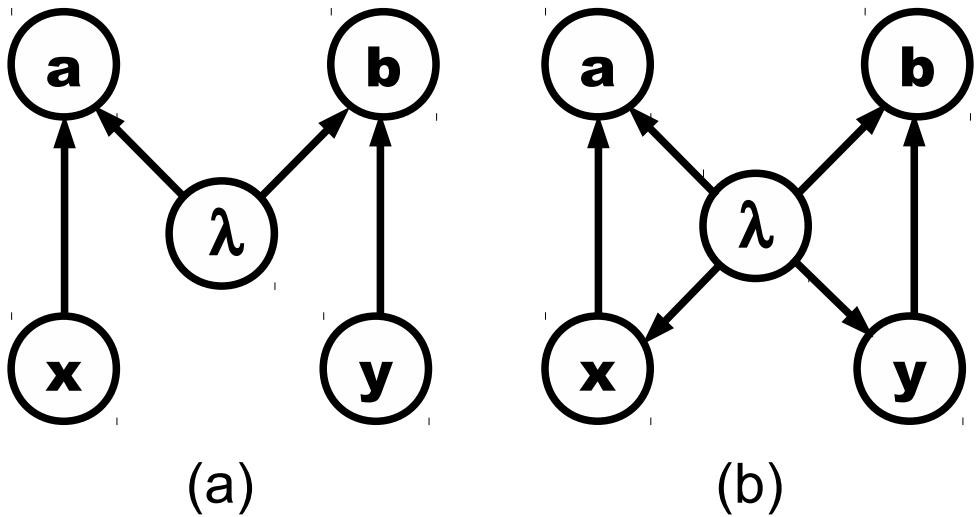}\\
  \caption{One hidden-variable model. The causal structure shows: (a) no causal connections between $\lambda$ and $x$ or $y$, therefore no violation of MIC (b) The hidden variable $\lambda$ is connected to $x$ and $y$, this is a violation of MIC}
\end{figure}

Bell's theorem present the possibility to test experimentally a theory satisfying three assumptions (here written in the language of causal model): \begin{enumerate}[I)] \item The value of the variable $a$ ($b$) is the join effect of a hidden variable $\lambda$ and the setting variable $x$ ($y$). \item The experiments performed by Alice and Bob are space-like separated events, therefore $a$ and $b$ are statistically independent given $\lambda$, $x$ and $y$. Or in mathematical terms $P(a,b|x,y,\lambda)=P(a|x,\lambda)P(b|y,\lambda)$. \item The measurement setting variables are independent of $\lambda$, that is the measurement independence condition. It can be written as $P(x,y|\lambda)=P(x,y)$ which is equivalent to $P(\lambda|x,y)=P(\lambda)$. \end{enumerate} Using the assumptions I, II and III, one can write the conditional probability $P(a,b|x,y)$ as:
\begin{eqnarray}
P(a,b|x,y)&=& \int d\lambda P(a,b|x,y,\lambda)P(\lambda|x,y),\notag\\
&=&\int d\lambda P(a|x,\lambda)P(b|y,\lambda)P(\lambda)\label{probell}.
\end{eqnarray} There is a conflict between the separable form of $P(a,b|x,y)$ in Equation \eqref{probell} and the predictions of quantum theory. This conflict is experimentally verified by violations of Bell inequality \cite{bell free}.

A singlet state (in the computational basis $|\psi_{singlet}\rangle = \frac{1}{\sqrt{2}}[|10\rangle - |01\rangle]$) is a maximally entangled state and can be used in experiments that show violation of a Bell inequality (for instance \cite{bell free}). Therefore, a model that satisfy assumptions I, II and III can not be used to predict the conditional probability $P_{S}(a,b|x,y)$ measured in a system prepared in a singlet state, which is \cite{art10}: 
\begin{equation}
P_{s}(a,b|x,y)=\frac{1-ab(x\cdot y)}{4}.
\end{equation}

If one of the assumptions (I, II or III) is relaxed one can reproduce, within a hidden variable model, the probability $P_{S}(a,b|x,y)$ of the singlet state. Here we consider relaxation on the measurement independence condition (MIC). Some models have been proposed \cite{art7,art8,art9,art10} in which the authors investigate the degree of relaxation of the MIC necessary to reproduce $P_{S}(a,b|x,y)$ for the singlet state. In the context of causal models, the relaxation of the MIC is represented by the causal connections between $\lambda$ and $x$, $y$ as shown in the directed acyclic graph (DAG) of Figure 1(b), i.e. there may be an explicit dependence of the measurement apparatus concerning the hidden variable. 

In the models presented in references \cite{art7,art8,art9,art10} the authors considered the relation $P(a,b|x,y)=\int d\lambda P(a,b|x,y,\lambda)P(\lambda|x,y)$ and chose suitable expressions for the conditional probabilities $P(a,b|x,y,\lambda)$ and $P(\lambda|x,y)$ to reproduce $P_{S}(a,b|x,y)$. The relation among the variables $\lambda,x,y$, determine the degrees of relaxation on the MIC. Therefore, each models have different degrees of relaxation on the MIC. The model with smaller degree of relaxation is the one of reference \cite{art10} (to our knowledge).

In the traditional Bell's scheme (Figure 1(a)), there is only one hidden variable $\lambda$ and a violation of the MIC appear when this variable has a causal link with $x$, $y$ as shown in Figure 1(b). If we consider causal models with more than one hidden variable, different causal links among $x$, $y$ and the hidden variables may be considered. Do these new structures allow us to reproduce the singlet probability $P_{S}(a,b|x,y)$ with a degree of relaxation smaller than the one in reference \cite{art10}? To partially answer this question we consider two models given in reference \cite{art11,art12}. The choice of this two models allow us to concentrate on the role played by the number of hidden variables.

\subsection{The Second Causal Model}

In this section we consider the model presented in \cite{art11}, whose causal structure is shown in Figure 2(a), and here we show how one can reproduce the singlet probability $P_{S}(a,b|x,y)$ whithin this model. Notice that the exogenous variables are the hidden variables, i.e. $\lambda_1, \lambda_2$ and $\lambda$. In the traditional Bell scheme -- Figure 1(a) -- the violation of the MIC is represented by the causal connections among the variables $x$, $y$ and $\lambda$, as it is shown in Figure 1(b). With the violation of the MIC, $x$ and $y$ cease to be exogenous variables, they become descendent \cite{pearl} of $\lambda$. In the model of Reference \cite{art11}, the violation of the MIC is represented by the causal connections among $\lambda$, $\lambda_{1}$ and $\lambda_{2}$, as shown in Figure 2(b), where $\lambda_{1}$ and $\lambda_{2}$ cease to be exogenous variables and become descendent of $\lambda$. 

\begin{figure}[h]
\centering
\hspace*{-0.7cm}
  \includegraphics[scale=0.4]{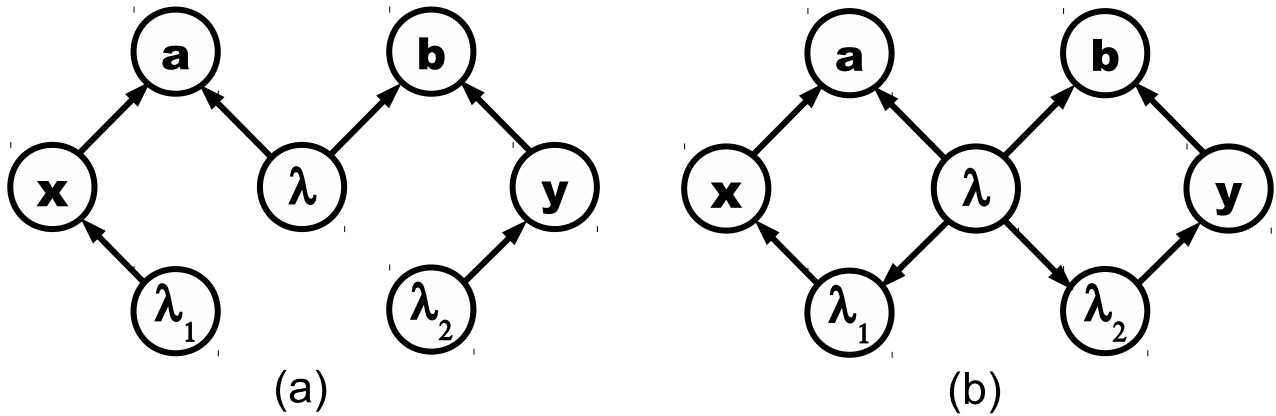}\\
  \caption{Three hidden-variable model. The causal structure shows: (a) no causal connections between $\lambda$ and $\lambda_{1}$ or $\lambda_{2}$, therefore no MIC violation; (b) now the hidden variable $\lambda$ is connected to $\lambda_{1}$ and $\lambda_{2}$, this is a clear MIC violation.}
\end{figure}

The conditional probability $P(a,b|x,y)$ is given by:

\begin{equation}
P(a,b|x,y)= \int\int\int d\lambda d\lambda_{1}d\lambda_{2} P(a,b|x,y,\lambda,\lambda_{1},\lambda_{2})P(\lambda,\lambda_{1},\lambda_{2}|x,y),\label{prob1}
\end{equation} from Bayes' theorem we can write \begin{equation}
P(\lambda,\lambda_{1},\lambda_{2}|xy) = \frac{P(x,y|\lambda,\lambda_{1},\lambda_{2})P(\lambda,\lambda_{1},\lambda_{2})}{P(x,y)}=\frac{P(x,y|\lambda,\lambda_{1},\lambda_{2})P(\lambda|\lambda_{1},\lambda_{2})P(\lambda_{1},\lambda_{2})}{P(x,y)}.\label{prob2}
\end{equation}

The causal Markov condition applied to the DAG of Figure 2(a) gives the relations: \begin{eqnarray}
P(a,b|x,y,\lambda,\lambda_{1},\lambda_{2})&=&P(a,b|x,y,\lambda)\label{prob3}\\
P(x,y|\lambda,\lambda_{1},\lambda_{2}) &=&P(x,y|\lambda_{1},\lambda_{2})\label{prob4}.  
\end{eqnarray}

Working out equations \eqref{prob1} to \eqref{prob4}, and using the definition $P(\lambda_{1},\lambda_{2}|x,y)=P(x,y|\lambda_{1},\lambda_{2})P(\lambda_{1},\lambda_{2})/P(x,y)$ we obtain:
\begin{equation}
P(ab|xy)= \int\int\int d\lambda d\lambda_{1}d\lambda_{2} P(a,b|x,y\lambda)P(\lambda_{1},\lambda_{2}|x,y)P(\lambda|\lambda_{1},\lambda_{2}).\label{modelo2}
\end{equation}

In the causal structure shown in Figure 2(a), variables $\lambda$, $\lambda_{1}$ and $\lambda_{2}$ are exogenous, therefore the causal Markov condition also return us the relation $P(\lambda|\lambda_{1},\lambda_{2})= P(\lambda)$. To investigate the relaxation of the MIC, let us consider the DAG shown in Figure 2(b). In this causal structure $\lambda_{1}$ and $\lambda_{2}$ are not exogenous and $P(\lambda|\lambda_{1},\lambda_{2})\neq P(\lambda)$.

In Reference \cite{art10} the author calculates, for the model in Figure 1(b), the degree of relaxation of the MIC necessary to reproduce the probability $P_{S}(a,b|x,y)$ of the singlet. In this work the author considers a particular relation among the variables $\lambda$, $x$ and $y$, and the degree of MIC obtained depends on this relation. To calculate the degree of relaxation for the second model and compare with the result obtained in Reference \cite{art10}, we consider the same relation among the exogenous variables, and substitute the measurement setting variables $x$ and $y$ by the hidden variables $\lambda_{1}$ and $\lambda_{2}$:

\begin{equation}
P(\lambda|\lambda_{1},\lambda_{2})=\frac{1}{4\pi}\cdot\frac{1+(\lambda_{1}.\lambda_{2})\sign[(\lambda .\lambda_{1})(\lambda .\lambda_{2})]}{1+(1-2\phi_{\lambda_{1} \lambda_{2}}/\pi)\sign[(\lambda .\lambda_{1})(\lambda .\lambda_{2})]},\label{modelo21}
\end{equation} where $\phi_{\lambda_{1} \lambda_{2}}$ represents the angle between the measurement
directions $\lambda_{1}$ and $\lambda_{2}$. To reproduce the singlet statistics we also consider the relations:
\begin{eqnarray}
P(a,b|\lambda,x,y)&=&\delta_{a,A(\lambda,x)}\delta_{b,B(\lambda,y)}\label{modelo22}\\
P(\lambda_{1},\lambda_{2}|x,y)&=&\delta_{\lambda_{1},x}\delta_{\lambda_{2},y}\label{modelo23}
\end{eqnarray} where $A(\lambda,x)=\sign(\lambda\cdot x)$ and $B(\lambda,y)=-\sign(\lambda\cdot y)$. Therefore, substituting Equations from \eqref{modelo21} to \eqref{modelo23} in Equation \eqref{modelo2} we obtain the singlet probability: \begin{equation}
P(a,b|x,y)=P_{S}(a,b|x,y).
\end{equation}

\subsection{The Third Causal Model}

In this section the causal model studied in reference \cite{art12} and shown in Figure 3(a) is considered. In this model, the condition equivalent to the measurement independence can be written as: $P(\lambda,\lambda_{1},\lambda_{2})=P(\lambda)P(\lambda_{1})P(\lambda_{2})$. 

\begin{figure}[h]
\centering
\hspace*{-0.7cm}
  \includegraphics[scale=0.4]{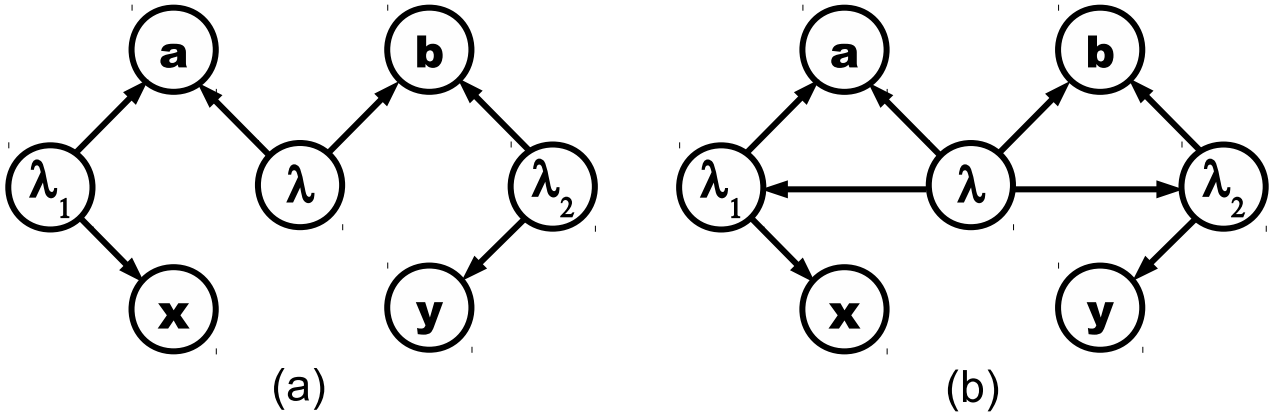}\\
  \caption{Three hidden-variable model. The causal structure shows: (a) no causal connections between $\lambda$ and $\lambda_{1}$ or $\lambda_{2}$, therefore no MIC violation; (b) the hidden variable $\lambda$ is connected to $\lambda_{1}$ and $\lambda_{2}$, a MIC violation.}
\end{figure}

To investigate the degree of relaxation necessary to reproduce the singlet statistics, we consider the causal structure given in Figure 3(b). The causal Markov condition permit us to write:
\begin{eqnarray}
P(a,b|x,y,\lambda,\lambda_{1},\lambda_{2})&=&P(a,b|\lambda,\lambda_{1},\lambda_{2})\label{prob5}\\
P(x,y|\lambda,\lambda_{1},\lambda_{2}) &=&P(x,y|\lambda_{1},\lambda_{2}).\label{prob6}
\end{eqnarray}

The conditional probability $P(a,b|x,y)$ can be written as:  
\begin{eqnarray}
P(a,b|x,y)&=& \int\int\int d\lambda d\lambda_{1}d\lambda_{2} P(a,b|x,y,\lambda,\lambda_{1},\lambda_{2})P(\lambda,\lambda_{1},\lambda_{2}|x,y),\nonumber \\
&=&\int\int\int d\lambda d\lambda_{1}d\lambda_{2} P(a,b|\lambda,\lambda_{1},\lambda_{2})P(\lambda,\lambda_{1},\lambda_{2}|x,y).\label{prob7}
\end{eqnarray}

Again from Bayes' theorem we can write:
\begin{equation}
P(\lambda,\lambda_{1},\lambda_{2}|x,y)= \frac{P(x,y|\lambda,\lambda_{1},\lambda_{2})P(\lambda,\lambda_{1},\lambda_{2})}{P(x,y)}.
\end{equation}

Using the definition of joint probability $P(\lambda,\lambda_{1},\lambda_{2})=P(\lambda|\lambda_{1},\lambda_{2})P(\lambda_{1},\lambda_{2})$ and Equation \eqref{prob6}, we can write:
\begin{equation}
P(\lambda,\lambda_{1},\lambda_{2}|x,y)= \frac{P(x,y|\lambda_{1},\lambda_{2})P(\lambda_{1},\lambda_{2})}{P(x,y)}P(\lambda|\lambda_{1},\lambda_{2})=P(\lambda_{1},\lambda_{2}|x,y)P(\lambda|\lambda_{1},\lambda_{2})\label{modelo3}
\end{equation}
 
In order to reproduce the probability of the singlet state $P_{S}(a,b|x,y)$, we use the strategy of the previous section. The variables involved in the MIC for this model are $\lambda$, $\lambda_{1}$ and $\lambda_{2}$, and we consider the same relation among them, substituting $x$ and $y$ by $\lambda_{1}$ and $\lambda_{2}$ (since $x$ and $y$ are not ascendant of any variable in this model, see Figure 3(b)), as it was done in the previous section. Then we obtain: \begin{equation}
P(\lambda|\lambda_{1}\lambda_{2})=\frac{1}{4}\frac{1+(\lambda_{1}.\lambda_{2})\sign[(\lambda .\lambda_{1})(\lambda .\lambda_{2})]}{1+(1-2\phi_{\lambda_{1} \lambda_{2}}/\pi)\sign[(\lambda .\lambda_{1})(\lambda .\lambda_{2})]}.\label{modelo31}
\end{equation}

To reproduce the singlet statistics we also consider the relations: \begin{eqnarray}
P(a,b|\lambda,\lambda_{1},\lambda_{2})&=&\delta_{a,A(\lambda,\lambda_{1})}\delta_{b,B(\lambda,\lambda_{2})}\label{modelo32}\\
P(\lambda_{1},\lambda_{2}|x,y)&=&\delta_{\lambda_{1},x}\delta_{\lambda_{2},y},\label{modelo33}
\end{eqnarray} where $A(\lambda,\lambda_{1})=\sign(\lambda\cdot \lambda_{1})$ and $B(\lambda,\lambda_{2})=-\sign(\lambda\cdot \lambda_{2})$. Again, working out Equations \eqref{modelo21} to \eqref{modelo23}, and substituting in Equation \eqref{modelo2}, we obtained the singlet probability:

\begin{equation}
P(a,b|x,y)=P_{S}(a,b|x,y).
\end{equation}

\section{Degree of Relaxation of Measurement Independence Condition}

In this section we compare the degree of relaxation of the MIC for the causal models represented in Fig. 1(b), Fig. 2(b) and Fig. 3(b). Some measures of the relaxation degree of the MIC have been considered in the literature \cite{art8,art10,art13,art14,art15,art16,art17,art18}, we follow \cite{art13} and use mutual information as our figure of merit. For the models shown in Figure 2(b) and in Figure 3(b) the violation of the MIC is due to the relations among variables $\lambda, \lambda_{1}$ and $\lambda_{2}$, therefore, the degree of violation of the MIC is given by the mutual information: 
\begin{equation}
I(\lambda_{1},\lambda_{2}:\lambda)=I(\lambda:\lambda_{1},\lambda_{2})= H(\lambda) - H(\lambda|\lambda_{1},\lambda_{2}),\label{ent}
\end{equation} where, $H(\phi)$ is the usual Shannon entropy related to some variable $\phi$. From Equation \eqref{ent} we can see that the degree of relaxation of MIC depends only on the conditional probability $p(\lambda|\lambda_{1},\lambda_{2})$ which are the same in the models shown in Fig 2(b) and Fig. 3(b). Therefore, the degree of relaxation will be the same for both models.

To compare the models with more than one hidden variable (Fig. 2(b) and Fig. 3(b)) with the traditional one (Fig. 1(b)), let us consider the difference $I(\lambda:\lambda_{1},\lambda_{2}) - I(\lambda:x,y)$, where $I(\lambda:x,y)$ represents the mutual information among the variables of interest in the model represented by Fig. 1(b). In this way, we obtained:
\begin{eqnarray}
I(\lambda:\lambda_{1},\lambda_{2}) - I(\lambda:x,y)&=& -H(\lambda|\lambda_{1},\lambda_{2})+H(\lambda|x,y)\label{dif}\\ 
&=&-\sum_{\lambda,\lambda_{1},\lambda_{2}}p(\lambda|\lambda_{1},\lambda_{2})\log\left[p(\lambda|\lambda_{1},\lambda_{2})\right]+\sum_{\lambda,x,y}p(\lambda|x,y)\log\left[p(\lambda|x,y)\right].\notag
\end{eqnarray}

Now we do need to make a digression and re-direct our attention back to the hidden variables domain, which we call $\Omega$. From Equations \eqref{modelo23} and \eqref{modelo33} we can conclude that, in order to reproduce the singlet statistics, $\Omega$ must contain $U$ (the set of unitary vectors and the domain of $x$ and $y$). Due to our lack of knowledge about the hidden variables, the cardinality of $\Omega$ is not known, but since one is interested in reproduce $P_{S}(a,b|x,y)$ within the causal models framework we are working, the cardinality of $\Omega$ must be greater than or equal to the cardinality of $U$. If the cardinalities are equal $I(\lambda:\lambda_{1},\lambda_{2}) - I(\lambda:x,y)=0$, but if they are different, Equation \eqref{dif} is non-zero:
\begin{equation}
I(\lambda:\lambda_{1},\lambda_{2}) - I(\lambda:x,y)= -\sum_{\lambda,\lambda_{1}\neq x,\lambda_{2}\neq y}p(\lambda|\lambda_{1},\lambda_{2})\log\left[p(\lambda|\lambda_{1},\lambda_{2})\right].
\end{equation}

As $p(\lambda|\lambda_{1},\lambda_{2})$ is a probability, we can write $0\leq p(\lambda|\lambda_{1},\lambda_{2})\leq 1$ and therefore $\log\left[p(\lambda|\lambda_{1},\lambda_{2})\right]\leq 0$. In conclusion, we obtain the inequality: 
\begin{eqnarray}
&&I(\lambda:\lambda_{1},\lambda_{2}) - I(\lambda:x,y) \geq 0\\
&&I(\lambda:\lambda_{1},\lambda_{2}) \geq I(\lambda:x,y). \label{ineq}
\end{eqnarray}

Inequality \eqref{ineq} shows that the degree of MIC violation, necessary to reproduce the singlet probability $P_{S}(a,b|x,y)$ within models of Fig. 2(b) and Fig. 3(b), is greater than or equal to the one calculated for the model of Reference \cite{art10}. 

\section{Conclusions} In this work we studied causal models where the measurement independence condition (as soon as we are dealing with hidden variables theories) may not be satisfied. Three causal models were studied, the first one  with one hidden variable, and the other two with 3 hidden variables. The model with one hidden variable was used to exemplify our approach and to obtain the probability distribution for the singlet state. In the following models we calculate the probability distribution, and we were able to obtain the statistics for the singlet state. Finally, we quantified the degree of relaxation for the studied cases. We show that the increase in the number of hidden variables, at least for the models studied in this work, does not allow the reduction of the mutual information needed to reproduce $P_{S}(a,b|x,y)$. 

\acknowledgments{The authors thanks Brazilian agencies CNPq and FAPEMIG for finantial support.}

\end{document}